# Is Inequality Among Universities Increasing?
# Gini Coefficients and the Elusive Rise of Elite Universities[*]

*Minerva* (forthcoming)

Willem Halffman [a] & Loet Leydesdorff [b]

[a] Science, Technology, and Policy Studies (STəPS), Twente University;
w.halffman@gmail.com

[b] Amsterdam School of Communications Research (ASCoR), University of Amsterdam;
loet@leydesdorff.net

**Abstract**
One of the unintended consequences of the New Public Management (NPM) in universities is often feared to be a division between elite institutions focused on research and large institutions with teaching missions. However, institutional isomorphisms provide counter-incentives. For example, university rankings focus on certain output parameters such as publications, but not on others (e.g., patents). In this study, we apply Gini coefficients to university rankings in order to assess whether universities are becoming more unequal, at the level of both the world and individual nations. Our results do not support the thesis that universities are becoming more unequal. If anything, we predominantly find homogenization, both at the level of the global comparisons and nationally. In a more restricted dataset (using only publications in the natural and life sciences), we find increasing inequality for those countries, which used NPM during the 1990s, but not during the 2000s. Our findings suggest that increased output steering from the policy side leads to a global conformation to performance standards.

## 1    Inequality among universities

Universities have increasingly been subject to output performance evaluations and ranking assessments (Frey & Osterloh, 2002; Osterloh & Frey, 2008). Performance indicators are no longer deployed only to assess university departments in the context of specific disciplines, but increasingly also to assess entire universities across disciplinary divides (Leydesdorff, 2008). Well-known examples are the annual Shanghai ranking, the *Times Higher Education Supplement* ranking, and the Leiden ranking, but governments also collect data at the national level about how their academic institutions perform.

        Not unlike restaurant or school ratings, university rankings convey the fascination of numbers despite the ambiguity of what is measured. A variety of interests convene around these numbers. Rankings seem to allow university managers to assess their organisation's performance, but also to advertise good results in order to attract additional resources. These extra resources can be better students, higher tuition fees, more productive researchers, additional funding, wider media exposure, or similar capital

[*] Acknowledgments: The authors wish to thank Heide Hackmann, Ronald Rousseau, and two anonymous reviewers for comments on a previous draft of this article.

increases. Rankings enable policy makers to assess national universities against international standards. Output indicators hold a promise of comparative performance measurement, suggesting opportunities to spur academic institutions to ever higher levels of production at ever reduced cost.

With university rankings, the competitive performance logic of New Public Management (NPM) further permeated into the academic sector (Martin, 2009; Shimank, 2005; Weingart & Maasen, 2007). The complex changes around NPM in the public sector involve a belief in privatisation (or contractual public-private partnerships) and quasi-market competition, an emphasis on efficiency and public service delivery with budgetary autonomy for service providers, with a shift from steering on (monetary) inputs to outputs, through key performance indicators and related audit practices (Power, 2005; Hood & Peters, 2004). In the academic sector, NPM has expressed itself with reduced state regulation and mistrust of academic self-governance, insisting instead on external guidance of universities through their clients, under a more managerial regime stressing competition for students and research resources – although the precise mix of changes varies between countries (De Boer, Enders & Schimank, 2007).

The expansion of performance measurement in the academic sector has incited substantial debate. Obvious objections concern the adequacy of the indicators. For example, the Shanghai ranking was criticised for failing to address varying publication levels among different research fields (Van Raan, 2005). In response to this critique, the methodology of the Shanghai ranking was adjusted: one currently doubles the number of publications in the social sciences in order to compensate for differences in output levels between the social and natural sciences. Going even further, the Leiden ranking attempts to fine-tune output measurement by comparing publication output with average outputs per field (Centre for Science and Technology Studies, 2008).[1]

In this article we focus on the debate about the consequences rather than methodology of output measurement. There is a growing body of research pointing to unwanted side-effects of counting publications and citations for performance measurement. Weingart (2005) has documented cases of ritual compliance, e.g., with journals attempting to boost impact factors with irrelevant citations. Similar effects are the splitting of articles to the 'smallest publishable unit' or the alleged tendency of researchers to shift to research that produces a steady stream of publishable data. Similar objections have been raised against other attempts to stimulate research performance through a few key performance indicators. Schmoch and Schubert (2009) showed that such a reduction may impede rather than stimulate excellency in research. As such, these objections are similar to objections voiced against NPM in other policy sectors, such as police organisations shifting attention to crimes with 'easy' output measurement, e.g. intercepted kilos of drugs, or schools grooming students to perform well on tests only. The debate over advantages and disadvantages of NPM is by no means closed (Hood & Peters, 2004).

One of the contested issues in the rise of NPM at universities is whether the new assessment regime would lead to increased inequality among universities (Van Parijs, 2009). According to the advocates of NPM, performance measurement spurs actors in the public sector into action. By making productivity visible, it becomes possible to compare performance and make actors aware of their performance levels. This can be expected to generate improvements, either merely through heightened awareness and a sense of obligation to improve performance, or through pressure from the actors' clients.

---

[1] The field normalization is based on using the ISI Subject Categories which are often unprecise and thus to be used only as a statistics (Rafols & Leydesdorff, 2009).

For example, by making the performance of schools visible, NPM claims that parents can make more informed choices about where to send their children. This transparency is expected to put pressure on under-performing schools. To stimulate actors even further, governments may tie the redistribution of resources to performance, as has been the case in the UK Research Assessment Exercises. The claim of NPM is that this stimulation of actors can be expected to improve the quality of public services and reduce costs. In the university sector, NPM promises more and better research at lower cost to the tax payer, in line with Adam Smith's belief in the virtues of the free market.

Opponents to the expansion of NPM into the university sector point to a number of objections that echo those made in other NPM-stricken public sectors. This is not the place to provide a complete overview of the debate; suffice it to say that the inequality in performance in the academic sector has been a crucial issue. While proponents of comparative performance measurement claim that all actors in the system will be stimulated to improve their performance, opponents claim that this ignores the redistributive effects of NPM. By moving university performance in the direction of commodification, NPM could create the accumulation of resources in an elite layer of universities, generating inequalities through processes that also produce the Matthew effect (Merton, 1968). These authors stress the down sides of the US Ivy League universities, including the creation of old boys' networks of graduates that produce an increasingly closed national elite, or the large inequalities of working conditions between elite and marginal universities.

In the same vein, critics claim that the aspirations of governments to have top-ranking universities, such as Cambridge or Harvard, may lead to the creation of large sets of insignificant academic organisations, teaching universities or professional colleges, at the other end of the distribution. In the case of Germany, where there has been much debate on inequalities among universities as a result of changes in academic policy, it has been argued that output evaluation practices reproduce status hierarchies between universities, affecting opportunities to attract resources (Münch, 2008).[2] In contrast to the belief in the general stimulation of actors, these critics appeal to a logic of resource concentration that is reminiscent of Marx's critique of oligopolistic capitalism.

A third and more constructivist understanding of performance measurement suggests that major shifts in the university sector cannot be expected to lead to an overall increase in performance, nor a shift of resources, but rather a widespread attempt of actors to 'perform performance'. If output is measured in terms of numbers of publications, then these numbers can be expected to increase, even at the expense of actual output: any activity that is not included in performance measurement will be abandoned in favour of producing good statistics. This reading of rankings considers them to be a force of performance homogenisation and control: a 'McDonaldisation of universities' (Ritzer, 1998), under a regime of 'discipline and publish' (Weingart & Maasen, 2007). These authors emphasize that the construction of academic actors who monitor themselves via output indicators may have even more detrimental effects than the capital destruction that comes with concentration. Output measurement is regarded as mutilating the very academic quality it claims to measure, through a process of Weberian rationalisation or an even more surreptitious expansion of governmentality, as signalled by Foucault (Foucault, 1991).

---

[2] An analysis of grants rewarded by the German science foundation showed no effect of institutional context on success of individual scientists' grant applications (Auspurg, Hinz & Güdler, 2008; cf. Van den Besselaar & Leydesdorff, 2009; Bornmann *et al.*, 2010).

Considering these serious potential consequences pointed out by the critics, there is surprisingly little systematic information on the changing inequalities among universities. Most of the debates rely on anecdotal evidence. Can one distinguish a top layer of increasingly elite universities that produce ever larger shares of science, at the expense of a dwindling tail of marginalized teaching universities? Ville *et al.* (2006) reported an opposite trend of equalization in research output among Australian universities (1992-2003) using Gini coefficients for the measurement. In this article, we use the Gini coefficient as an indicator for assessing the development of inequalities in academic output in terms of publications at the global level. The Gini measure of inequality is commonly used for the measurement of income inequalities and has intensively been used in scientometric research for the measurement of increasing (or decreasing) (in)equality (e.g., Bornmann *et al*., 2008; Cole *et al*., 1978; Danell, 2000; Frame *et al*., 1977; Persson & Melin, 1996; Stiftel *et al*., 2004; Zitt *et al*., 1999). Burrell (e.g., 1991) and Rousseau (e.g., 1992, 2001), among others, studied the properties of Gini in the bibliometric environment (cf. Atkinson, 1970).

By providing a more systematic look at the distribution of publication outputs of universities and the potential shifts of these distributions over time, we hope to contribute with empirical data to the ongoing debate of the merits and drawbacks of comparative performance measurement in the university sector. Although we use indicators such as the Shanghai ranking or output measures in this article, we do not consider these to be unproblematic or desirable indicators of research performance. Rather, we want to investigate how the distribution of outputs between universities changes, irrespective of what these outputs represent in terms of the 'quality' of the universities under study. This implies that we do not want to take sides in the debate on the value of output measurement, but rather test the claims that are made about the effects of NPM in terms of the outputs it claims to stimulate. Which version is more plausible: the NPM argument of stimulated performance in line with Adam Smith, the fear of increasing elitism reminiscent of Marx' logic of capital concentration, or the constructivist reading following Foucault's spread of governmentality and discipline?

## 2   Methods and data

The Gini indicator is a measure of inequality in a distribution. It is commonly used to assess income inequalities of inhabitants or families in a country. Gini indicators play an important role in the redistributive policies of welfare states, e.g., to assess whether all layers of the population share in collective wealth increases (Timothy, 2005). They also play a key role in the debate about whether or not global inequalities are increasing (Dowrick & Akmai, 2006; Sala-i-Martin, 2006). In the case of income distributions, the Ginis of most Northern European countries are around 0.25 (Netherlands, Germany, Norway), while the Gini coefficient of the USA is 0.37. For Mexico—as an example of the relatively unequal countries in Latin America—the Gini coefficient is 0.47 (Timothy, 2005).

In order to calculate the Gini indicator, one orders the units of analysis—in our case, universities—from the lowest to highest output and plots a curve that shows the cumulative output: the first point in the plot corresponds to output of the smallest unit in these terms, the next is the smallest plus the one-but smallest, etc. This leads to the so-called Lorenz curve. In a perfectly 'equal' system, all universities would contribute the same share to the overall output. In that case, the Lorenz curve would be a straight line. In the most extremely unequal system, all universities but one would produce zero publications. A single university would produce all publications in the system, and the Lorenz curve would follow the *x*-axis until this last point is reached.

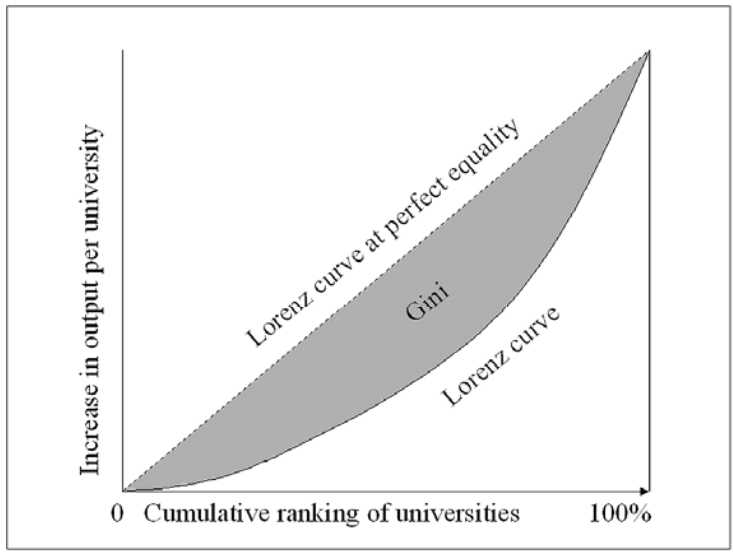

**Figure 1 Lorenz curve and Gini coefficient.**

Based on this reasoning, the Gini coefficient measures the relative surface between the Lorenz curve and the straight line (Figure 1). The Gini coefficient can be formulated as follows (Buchan, 2002):

$$G = \frac{\sum_{i=1}^{n}(2i-n-1)x_i}{n\sum_{i=1}^{n}x_i} \qquad (1)$$

with *n* being the number of universities in the population and $x_i$ being the number of publications of the university with position *i* in the ranking. Hence, the Gini ranges between zero for a completely equal distribution and $(n-1)/n$ for a completely unequal distribution, approaching one for large populations. For comparison among smaller populations of varying size, this requires a normalisation that brings Gini coefficients for all populations to the same maximum unity. The formula for this normalised Gini coefficient is:

$$G_N = \frac{n}{(n-1)} \frac{\sum_{i=1}^{n}(2i-n-1)x_i}{n\sum_{i=1}^{n}x_i} = \frac{\sum_{i=1}^{n}(2i-n-1)x_i}{(n-1)\sum_{i=1}^{n}x_i} \qquad (2)$$

Although statistical in nature, the Gini index is a relatively simple and robust measure of inequality. However, there are some complications. First, the Gini coefficient is sensitive to tails at the top or bottom of the distribution. At the top end, the inclusion or omission of one more highly productive university would alter the Gini drastically. In our data, however, these top-universities are also the most visible ones (e.g., Harvard, Oxford, Tokyo) and hence such an omission is unlikely in this study. At the bottom of the range, the data contains long tails of universities with very small numbers of publications; relatively unknown institutions, often even hard to recognise as universities. This problem can be resolved by comparing only fixed ranges, for example, the top-500 most productive universities. For the world's leading scientific countries this makes little difference. For example, our counts for the Shanghai ranking systematically include 12 of the 14 Dutch universities, 40 universities of some 120 universities in the UK and 159 of some 2000 universities and colleges for the USA. Nevertheless, this admittedly does exclude the very bottom of the range, and it may have an effect when we compare over time, as we shall see below for the case of China.

A second complication arises from double counts or alternate names of universities. For example, publications may be labelled as university or university medical centre publications; universities may change names over time, merge, or split. All of this creates larger or smaller units that will alter the distribution and hence the Gini. Therefore, it is important that publication data are carefully labelled, or at least consistently labelled over time. This requires a manual check.

Third, the Gini remains only a measure of overall inequality. This facilitates comparison from year to year, but the measure does not allow us to locate where changes in the distribution occur. To this end, Gini analysis can be complemented with comparisons of subset shares in overall output, such as the publication share of the top quartile or decile (10%) (cf. Plomp, 1990).

In order to calculate inequality among universities, we have first used the university output data provided by the Shanghai rankings at http://www.arwu.org. These rankings consist of a compounded indicator, with weighted contributions of total numbers of publications per university, awards won by employees of the university and alumni, and publications per researcher, in addition to numbers of highly cited publications and publications in *Nature* and *Science* by the top universities' scientists. For presentation purposes, the ranking scores of universities are expressed as a percentage of the top university (Harvard), but for the calculation of Gini-coefficients this normalization does not make a difference.

The central part of the Shanghai ranking only pertains to the world's top-50 universities, but publication data is provided for a larger set of 500 universities covering the years 2003-2008. The data other than numbers of publications for these top-500 is problematic because of cumulative scoring over years (e.g., for awards) or shifts in the data definition (e.g. inclusion of Fields awards in addition to Nobel Prizes). Unfortunately, the number of publications per scientist has also been adjusted during the series. The relevant definition is stable for the period 2005-2008.

Although this data provides us with a solid base for measuring inequalities, the time series is very short. For the precise ranking of each individual university in each year, the precision of total publications as a measure of productivity may be problematic. For our purposes, however, it makes little difference whether a specific university of—say, Manchester—follows at position number 40 (in 2008) or 48 (in 2007). The focus is on the shape of the distribution.

In order to investigate longer-term trends, additional calculations were performed on *Science Citation Index* data. Our data comprise results for the natural sciences only, but allow us to analyse developments over a longer period (1990-2007). Following best practice in scientometrics, we used only citable items, that is, articles, reviews, and letters.[3] More than 60% of the addresses are single occurrences; these include also addresses with typos. Using only the institutional addresses which occurred more than once—21,393 in 1990, but 46,339 in 2007—we removed all non-university organisations from the list and merged alternate names of the same universities. We included academic hospitals as separate organisations as part of our effort to limit manual intervention in the data to a minimum. For the analysis of shifts in the distribution over time, we believe that consistency is more important than debatable re-categorisations.

We should stress that our parameter, total SCI publications, can as much be considered as an indicator of size as of productivity. For example, at the top of our list is not Harvard, but the much larger University of Texas (see Table 1 for the top-50 largest universities in 2007). When we talk about the largest or the top universities, we refer to this measure of total SCI-covered publication output. We cannot make any claims about the long tail of small universities, but our analysis reaches as far down as Hunan University (532 SCI publications in 2007), St Louis (540 publications), or Bath (588 publications).

| Institute | Total | Country |
|---|---|---|
| Univ Texas | 12047 | USA |
| Harvard Univ | 11479 | USA |
| Univ Tokyo | 7435 | JAPAN |
| Univ Toronto | 7120 | CANADA |
| Univ Calif Los Angeles | 6803 | USA |
| Univ Michigan | 6603 | USA |
| Univ Washington | 6348 | USA |
| Univ Illinois | 5630 | USA |
| Kyoto Univ | 5465 | JAPAN |
| Johns Hopkins Univ | 5455 | USA |

---

[3] On February 27, 2009, Thomson-Reuters ISI announced a reorganization of the database in October 2008 (at http://isiwebofknowledge.com/products_tools/multidisciplinary/webofscience/cpci/usingproceedings/; Retrieved on March 11, 2009) . An additional category of citable "Proceedings Papers" is now distinguished on the Web-of-Science. Our data is not affected by this change since based on the CD-Rom versions of the *Science Citation Index.*

| Stanford Univ | 5447 | USA |
| --- | --- | --- |
| Univ Pittsburgh | 5442 | USA |
| Univ Wisconsin | 5369 | USA |
| Univ Penn | 4977 | USA |
| Univ Calif San Francisco | 4962 | USA |
| Univ Calif Berkeley | 4956 | USA |
| Univ Calif San Diego | 4942 | USA |
| Univ Minnesota | 4742 | USA |
| Seoul Natl Univ | 4687 | SOUTH KOREA |
| Columbia Univ | 4645 | USA |
| Univ Sao Paulo | 4628 | BRAZIL |
| Duke Univ | 4587 | USA |
| Tohoku Univ | 4579 | JAPAN |
| Univ Florida | 4450 | USA |
| Osaka Univ | 4433 | JAPAN |
| Univ N Carolina | 4406 | USA |
| Univ Calif Davis | 4379 | USA |
| Ohio State Univ | 4342 | USA |
| Univ Maryland | 4283 | USA |
| Yale Univ | 4195 | USA |
| Univ British Columbia | 4094 | CANADA |
| Mcgill Univ | 4048 | CANADA |
| Washington Univ | 4036 | USA |
| Cornell Univ | 4028 | USA |
| Univ Cambridge | 4018 | ENGLAND |
| Univ Colorado | 4007 | USA |
| Univ Oxford | 3879 | ENGLAND |
| Mit | 3850 | USA |
| Natl Taiwan Univ | 3848 | TAIWAN |
| Penn State Univ | 3654 | USA |
| Northwestern Univ | 3621 | USA |
| Univ Helsinki | 3515 | FINLAND |
| Vanderbilt Univ | 3398 | USA |
| Natl Univ Singapore | 3348 | SINGAPORE |
| Univ Paris 06 | 3289 | FRANCE |
| Univ Coll London | 3255 | ENGLAND |
| Zhejiang Univ | 3203 | PEOPLES R CHINA |
| Univ Alabama | 3193 | USA |
| Univ Sydney | 3184 | AUSTRALIA |
| Univ Melbourne | 3170 | AUSTRALIA |

**Table 1 The 50 largest universities in the world in 2007, in terms of totals of SCI publications.**

# 3 Results

## 3.1 *Inequality among the top-500 universities: Shanghai ranking data*

Gini coefficients for university publication output, based on the Shanghai ranking data, seem to remain stable between 2003 and 2008 (Figure 2). If anything, the overall inequality among universities decreases slightly. In any case, there is no indication of a significant and lasting increase in inequality as predicted on the basis of qualitative observations (e.g., Martin, 2009; Van Parijs, 2009, at p. 203).

|  | 2003 | 2004 | 2005 | 2006 | 2007 | 2008 | Avg 03-08 | *n* |
|---|---|---|---|---|---|---|---|---|
| World | 0.195 | 0.196 | 0.196 | 0.195 | 0.188 | 0.187 | 0.193 | 500 |
| Australia | 0.191 | 0.187 | 0.184 | 0.196 | 0.198 | 0.195 | 0.192 | 13 |
| Canada | 0.175 | 0.175 | 0.166 | 0.171 | 0.169 | 0.174 | 0.172 | 21 |
| China | 0.106 | 0.108 | 0.108 | 0.098 | 0.082 | 0.084 | 0.098 | 8 |
| France | 0.190 | 0.187 | 0.209 | 0.199 | 0.166 | 0.179 | 0.188 | 21 |
| Germany | 0.099 | 0.119 | 0.120 | 0.120 | 0.118 | 0.121 | 0.116 | 40 |
| Italy | 0.141 | 0.143 | 0.146 | 0.147 | 0.183 | 0.143 | 0.150 | 20 |
| Japan | 0.223 | 0.219 | 0.229 | 0.237 | 0.227 | 0.236 | 0.228 | 31 |
| Netherlands | 0.126 | 0.127 | 0.129 | 0.120 | 0.124 | 0.119 | 0.124 | 12 |
| Sweden | 0.122 | 0.120 | 0.121 | 0.121 | 0.132 | 0.134 | 0.125 | 10 |
| UK | 0.187 | 0.198 | 0.194 | 0.185 | 0.184 | 0.189 | 0.190 | 40 |
| US | 0.222 | 0.214 | 0.211 | 0.209 | 0.212 | 0.215 | 0.214 | 159 |

**Table 2 Normalised Gini coefficients for university publication outputs. Source: Shanghai ranking data at http://www.arwu.org/.**

Figure 2 shows remarkable differences in inequality among national systems. Here, we have to proceed with some caution, as the bottom tail of least productive universities may not be included to the same extent for all nations. China, for example, presents a problem, because ten more universities entered the top-500 between 2003 and 2008. All our calculations were made with the largest available set for all the years involved. (Hence, *n* is the same for every year.)

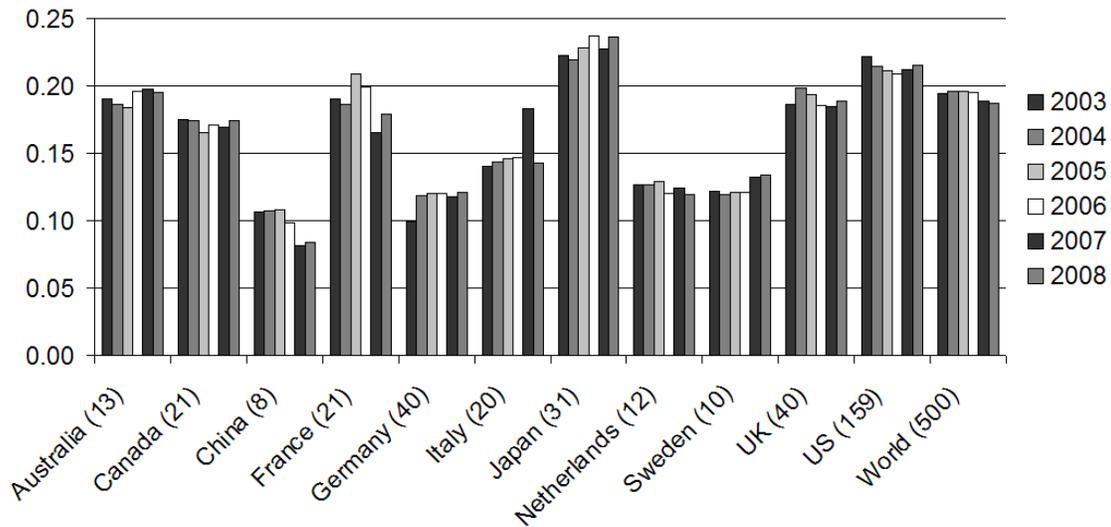

**Figure 2 Normalised Gini coefficients for university publication output. Source: Shanghai ranking at http://www.arwu.org/.**

Figure 2 shows a relative equality in the university systems of the Netherlands, Sweden, and Germany. We must point out that this does not mean that all universities in the respective countries are equally 'good', but rather that these universities produce a relatively similar number of publications. Inversely, the relatively high inequalities in Japan, the UK, or the US could just as well be caused by large differences in the size of universities as of their productivity.

Perhaps more remarkably, we do not observe major shifts in inequality over time within each national system. This is especially interesting for countries such as the UK, where increased inequalities could have occurred due to the redistribution effects of the Research Assessment Exercises. These research assessments redistribute research resources to the more productive research units, while reducing the budgets of those that do poorly in the evaluations. France and Italy, both in the middle range, display one or two erratic results, which we fear may be due to data redefinitions.

The lack of clear-cut increases in inequality among universities in terms of publication output raises further questions about productivity. What is happening to the outputs of publications per scientist? Because the use of the Gini coefficient is questionable here, as productivity data cannot be added meaningfully, we have used a simple standard deviation to measure dispersion. This is not quite the same as inequality, but does provide an indication of changes in the spread of productivity. The data is more irregular here, due to adjustments and improvements in the ranking data from year to year (Figure 3). Here too, one sees no clear sign of growing disparities among universities. The world trend seems slightly in favour of increasingly similar output levels (Leydesdorff & Wagner, 2009). Once again, the US ranks high in terms of spread in productivity levels, but Japan is now a member of the middle range. This implies that Japan may have a relatively large disparity between larger and smaller universities, but with more equal productivity levels. In the case of Australia, this difference is even larger, with the most equal distribution of productivity (Sdev = 3.7) among the other countries analysed, not considering China.

|           | 2005 | 2006 | 2007 | 2008 | Avg 05-08 | n   |
|-----------|------|------|------|------|-----------|-----|
| World     | 9.4  | 9.3  | 9.2  | 9.4  | 9.3       | 500 |
| Australia | 3.6  | 3.8  | 3.8  | 3.4  | 3.7       | 13  |
| Canada    | 7.3  | 8.0  | 7.9  | 8.1  | 7.8       | 21  |
| China     | 3.0  | 3.5  | 2.2  | 2.3  | 2.7       | 8   |
| France    | 5.3  | 5.8  | 6.1  | 9.9  | 6.8       | 21  |
| Germany   | 4.9  | 5.2  | 5.1  | 5.2  | 5.1       | 40  |
| Italy     | 7.9  | 7.8  | 7.9  | 6.0  | 7.4       | 20  |
| Japan     | 6.2  | 6.2  | 5.9  | 6.2  | 6.1       | 31  |
| Netherlands | 4.0 | 4.1 | 4.2 | 4.3 | 4.2       | 12  |
| Sweden    | 5.1  | 5.1  | 5.0  | 5.1  | 5.1       | 10  |
| UK        | 11.5 | 8.3  | 8.2  | 8.0  | 9.0       | 40  |
| US        | 12.3 | 12.3 | 12.3 | 12.4 | 12.3      | 159 |

**Table 3 Standard deviations for publication output per scientist. Source: Shanghai ranking data at http://www.arwu.org/.**

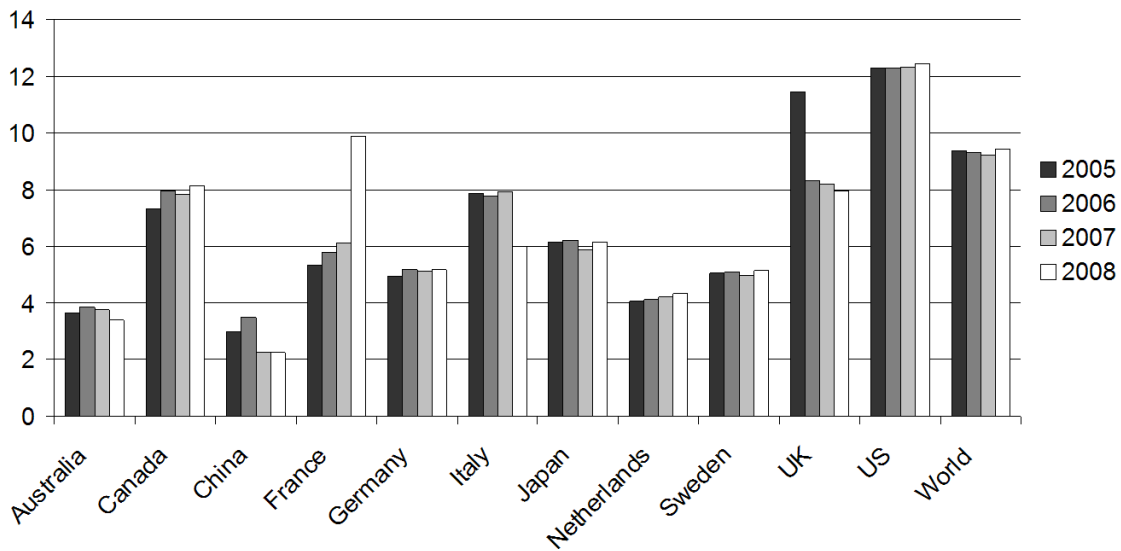

**Figure 3 Standard deviations for top-500 universities: productivity in SCI publications per faculty. Source: Shanghai ranking data at http://www.arwu.org/.**

Our results undermine the hypothesis of increasing inequalities among universities. If anything, we see a small *decrease* in output inequalities among universities, in terms of both overall output and productivity. This raises additional questions. Is this result the product of the methodological flaws of the Shanghai ranking (Van Raan, 2005), even if one uses only its least problematic component, that is, publication data derived from the *Science Citation Index*? Might we have missed the increasing formation of super-universities because the time frame used was too narrow? In order to answer these longitudinal questions, we turned to data sets from the *Science Citation Index* (*SCI*) for earlier years.

## *3.2 Inequality between universities: SCI data*

The 500 universities that publish most in the world, using the *SCI*, are becoming more equal in terms of their publication output. The trend is clear from 1990 to 2005 and continues thereafter for 2006 and 2007, confirming what we have found on the basis of the Shanghai ranking for a shorter time span (Figure 4). The relative position of the countries is similar to that in the Shanghai ranking, also confirming the measurement.

The trend per country shows a somewhat different picture. In the UK, the US, the Netherlands, Canada, and Australia, we see increases in inequality between 1990 and 2005, although these seem to decrease for the first three of these countries during recent years. These are also the countries in which NPM has been picked up early. However, whereas the UK has attached a redistribution of resources to research assessment, other countries, such as the Netherlands, have not.

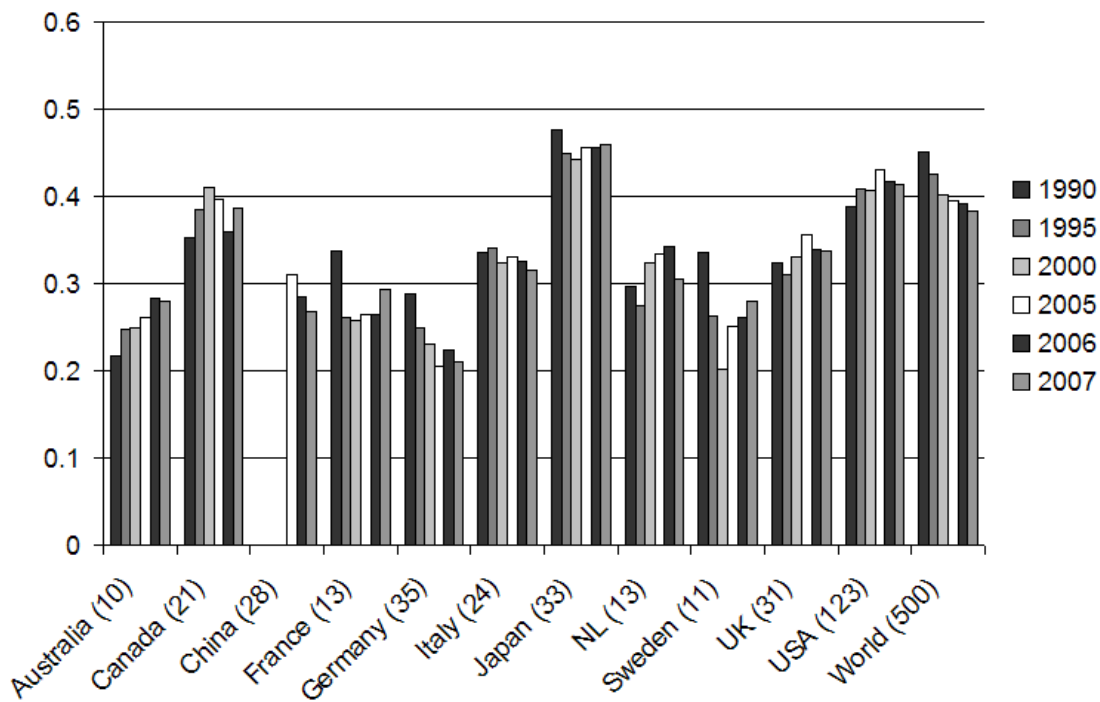

**Figure 4 Normalized Gini coefficients for top-500 universities. Source: SCI, *n* of publications (in brackets).**[4]

---

[4] The requirement to keep the number of universities per countries stable in order to calculate a comparable national Gini coefficient across the years led in the case of China to using a cut-off point of 28 universities

France, Italy, and Japan show a stable distribution of outputs, while there is a trend toward more equality in China, Germany, and Sweden, although with some erratic movement in the latter case. Although the overall image is consistent with the above results using data from the Shanghai rankings, the country patterns are different. However, these differences in trends are mainly the result of the expanded time horizon. For recent years at least the direction of the country trends is consistent with the Shanghai findings. Note that in all cases, the inequalities measured in the SCI are considerably larger than using the Shanghai ranking, which suggests that the natural sciences are more unequally distributed than the social sciences because the latter are included in the Shanghai ranking and not in the SCI data.

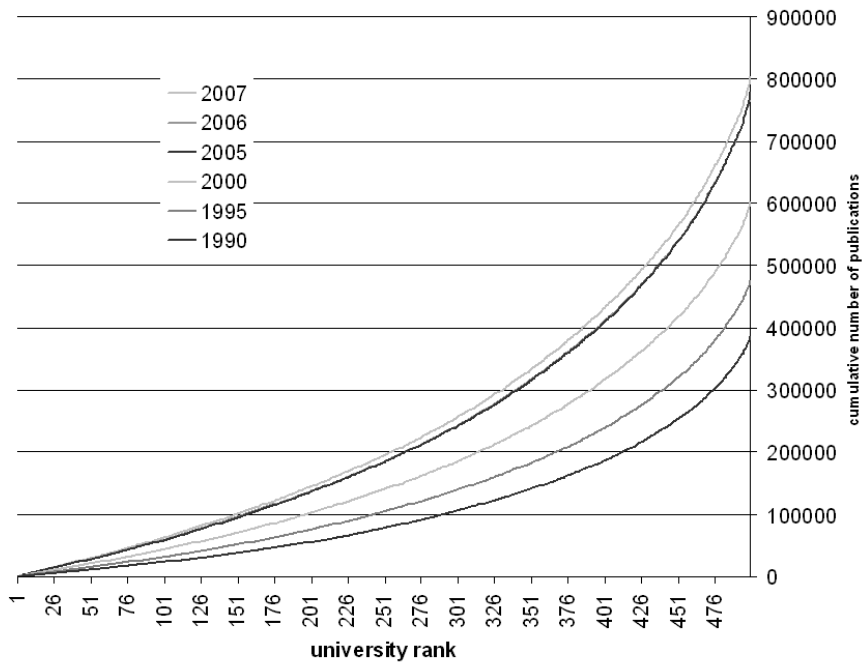

**Figure 5 Lorenz curves SCI publications 500 largest universities. Source: SCI.**

The Lorenz curves (Figure 5) show first the expansion of the database during the period under study. The 500 largest universities have increased their numbers of SCI publications, accordingly, from just under 400,000 in 1990 to almost 800,000 publications per year in 2007. This figure provides us with an impression of the evolution of the distribution, but in order to obtain a more precise understanding, we need to analyse the distributions in more detail.

## 3.3  *Details of the distribution*

Since much of the policy debate around rankings concerns aspirations to perform like the international top-universities, it is interesting to look in more detail at what the largest

---

in the years 2005-2007, disregarding the earlier presence of three Chinese universities among the top-500 in 1990, five in 1995, and sixteen in 2000.

universities are doing. To this end, we analysed the shares of total publications produced by every quarter, every tenth (decile), and every hundredth section of the distribution. We report the deciles here, as they provide the clearest indication of where the distribution is shifting (Table 4).

The top decile of universities is very slowly but steadily loosing ground in terms of output share. Whereas the 50 largest universities produced 34.4% of all SCI publications in the world in 1990, this share had decreased to 30.3% in 2007. This is not exactly a landslide, but in any case not an indication of a stronger oligopolistic concentration. Combined, the bottom half of the distribution has increased its share from a fifth (20.6%) to almost a quarter (24.0%) of the top-500 output (Figure 6).

|  | 1990 | 1995 | 2000 | 2005 | 2006 | 2007 | 2007-1990 |
|---|---|---|---|---|---|---|---|
| D1 | 2.9% | 2.9% | 3.2% | 3.4% | 3.5% | 3.5% | 0.7% |
| D2 | 3.3% | 3.4% | 3.7% | 3.9% | 4.0% | 4.1% | 0.8% |
| D3 | 3.9% | 4.0% | 4.4% | 4.6% | 4.6% | 4.7% | 0.8% |
| D4 | 4.7% | 4.9% | 5.3% | 5.3% | 5.3% | 5.4% | 0.7% |
| D5 | 5.9% | 6.0% | 6.2% | 6.4% | 6.2% | 6.3% | 0.4% |
| D6 | 7.3% | 7.3% | 7.6% | 7.6% | 7.5% | 7.7% | 0.4% |
| D7 | 9.0% | 9.2% | 9.5% | 9.6% | 9.4% | 9.5% | 0.5% |
| D8 | 11.8% | 12.1% | 12.4% | 12.1% | 12.2% | 12.2% | 0.4% |
| D9 | 16.9% | 17.0% | 16.6% | 16.4% | 16.6% | 16.2% | -0.6% |
| D10 | 34.4% | 33.1% | 31.1% | 30.8% | 30.8% | 30.3% | -4.1% |
| Tot | 100.0% | 100.0% | 100.0% | 100.0% | 100.0% | 100.0% |  |

**Table 4 Decile shares of the top-500 universities. Source: SCI.**

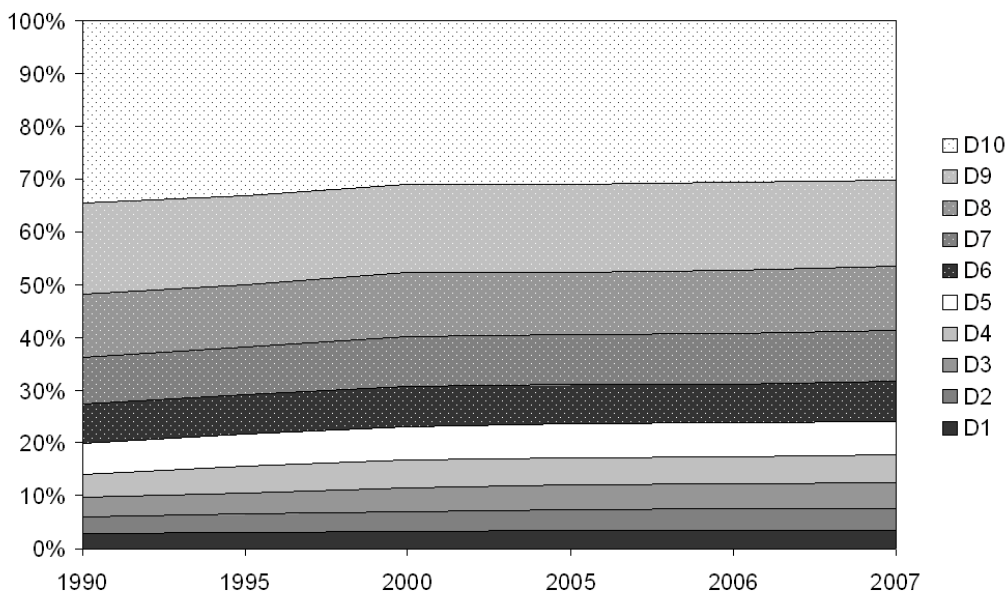

**Figure 6 Cumulative decile shares in total SCI output top-500 universities.**

A detailed analysis of the top ten percentiles showed that the decreasing share of the top decile was shared throughout the fifty largest universities and strongest in the top percentiles. Among the 100 largest universities, the Gini coefficient has <u>decreased</u> from 0.230 to 0.211 between 1990 and 2007.

## 4   Conclusion

Our results suggest an ongoing homogenization in terms of publication and productivity patterns among the top-500 universities in the world. Especially, the fifty largest universities are slowly loosing ground, while the lowest half of the top-500 catches up. All of this occurs against the background of rising output in all sections, and further expansion of the ISI-databases. In summary, it appears that the gap between the largest universities and the rest is closing rather than widening. Note that the top-500 universities are concentrated in North America, Western Europe and some Asian countries (Leydesdorff & Wagner, 2008). Within this set, we found increasing inequality in some countries between 1990 and 2005 using the SCI data, notably in the Anglo-Saxon world. However, even in these countries the trend seems to reverse in more recent years. Using a similar methodology, Ville *et al.* (2006) found decreasing inequality in research outputs among Australian universities during the period 1992-2003 given relatively stable funding distributions within this country.

In terms of Marxist, neo-liberal, and Foucauldian accounts of NPM, these results seem to refute the thesis suggesting oligopolistic tendencies in the university system, at least in terms of output. Further studies would have to analyse whether this trend is also present in the inputs of universities, such as research budgets, number of faculty, or even tuition fees. The Matthew effect, which generates concentration of reputation and resources in the case of individual scientists, if at all at work at the meso level of organisations, may have generated inequalities among universities in the past, but this process seems to have reached its limit. Perhaps the largest universities are now also facing disadvantages of scale.

The question remains whether the slow levelling off corroborates the idea that the neo-liberal logic of activation is responsible for this result, or whether the Foucauldian reading carries more weight. There are indications that universities are indeed shifting their output more towards what is valued in the rankings and output indicators such as SCI publications. Leydesdorff and Meyer (2010) have observed that the increase in publication output may be achieved at the expense of patents output since approximately 2000. The prevailing trend of levelling off of productivity differences in recent years also suggests that universities worldwide are conforming to isomorphic pressures of producing the same levels of SCI outputs. This further suggests that the self-monitoring of research actors increasingly follows the same global standards (DiMaggio & Powell, 1983).

There may be a price to pay for such higher output levels, apart from the family life of researchers. In the Netherlands, one witnesses a devaluation of publications in national journals for the social sciences, to the extent that several Dutch social science journals have recently ceased to exist because of a lack of good copy. Such trends have been criticised for undermining the contributions that the social sciences and humanities can make to national debates and public thought (Boomkens, 2008). Anecdotal evidence further suggests that researchers consciously shift to activities that produce a regular stream of publications, or that research evaluations may favour such research lines (Weingart, 2005; Laudel & Orrigi, 2006). Such evidence suggests that the slow levelling off of scientific output may not support the neo-liberal argument for increased competition at all. Rather, it suggests that researchers become better at 'performing performance,' i.e., the ritual production of output in order to score on performance indicators, even at the expense of the quality of one's work. Further research about the effects of NPM on universities will have to provide more clarity on these issues. Hitherto, the NPM wave has been programmatically resilient against counter-indications such as unintended consequences (Hood & Peters, 2004).

Whereas the inequality of scientific production has received scholarly attention in the past (Merton, 1968; Price, 1976), this discussion has focused mainly on the dynamics of reward structures of individuals and departments (Whitley, 1984). However, inequality at the institutional level of universities remains topical in the light of the NPM discussion (Martin, 2009). Our findings suggest that increased output steering from the policy side leads to a global conformity to performance standards, and thus tends to have unexpectedly an equalizing effect. Whether countries adopt NPM or other regimes to promote publication behavior (e.g., China) does not seem to play a crucial role in these dynamics.

**References**

Atkinson, A. B. (1970). On the measurement of inequality, *Journal of Economic Theory*, 2, 244-263

Auspurg, K., Hinz, T., & Gudler, J. (2009). Emergence of an Academic Elite? *Kölner Zeitschrift für Sozialpsychologie*, 60, 653-685.

De Boer, H., Enders, J., & Schimank, U. (2007). On the way towards new public management? The givernance of university systems in England, the Netherlands, Austria, and Germany. In D. Jansen (Ed.), *New Forms of Governance in Research Organizations. Disciplinary Approaches, Interfaces and Integration* (pp. 137-152). Dordrecht: Springer.

Boomkens, R. (2008). *Topkitsch en slow science*. Amsterdam: Van Gennep.

Bornmann, L., Leydesdorff, L., & Van den Besselaar, P. (2010). A Meta-evaluation of Scientific Research Proposals: Different Ways of Comparing Rejected to Awarded Applications. *Journal of Informatics,* in print; doi:10.1016/j.joi.2009.10.004

Bornmann, L., Mutz, R., Neuhaus, C., & Daniel, H. D. (2008). Citation counts for research evaluation: standards of good practice for analyzing bibliometric data and presenting and interpreting results. *Ethics in Science and Environmental Politics(ESEP),* 8(1), 93-102.

Buchan, I. (2002). Calculating the Gini coefficient of inequality. Retrieved on 11 March, 2009, from https://www.nibhi.org.uk/Training/Forms/AllItems.aspx?RootFolder=%2FTraining%2FStatistics&FolderCTID=&View={4223A4850-B4790-4965-4285DB-D4220A4841A5430B}.

Burrell, Q. L. (1991). The Bradford distribution and the Gini index. *Scientometrics,* 21(2), 181-194.

Centre for Science and Technology Studies. (2008). The Leiden Ranking 2008. Retrieved on 9 March, 2009, from http://www.cwts.nl/ranking/LeidenRankingWebSite.html.

Cole, S., Cole, J., & Dietrich, L. (1978). Measuring the Cognitive State of Scientific Disciplines. In Y. Elkana, J. Lederberg, R. K. Merton, A. Thackray & H. Zuckermann (Eds.), *Toward a Metric of Science: The Advent of Science Indicators* (pp. 209-251). New York, etc.: Wiley.

Danell, R. (2000). Stratification among journals in management research: a bibliometric study of interaction between European and American journals. *Scientometrics,* 49(1), 23-38.

DiMaggio, P. J., & Powell, W. W. (1983). The Iron Cage Revisited: Institutional isomorphism and collective rationality in organizational fields. *American Sociological Review,* 48, 147-160.

Dowrick, S., & Akmai, M. (2006). Contradictory trends in global income inequality: a tale of two biases. *The Review of Income and Wealth,* 51(2), 201-229.

Foucault, M. (1991). Governmentality. In G. Burchell, C. Gordon & P. Miller (Eds.), *The Foucault effect: Studies in Governmentality: with Two Lectures by and an Interview with Michel Foucault* (pp. 87-104). Chicago: University of Chicago Press.

Frame, J. D., Narin, F., & Carpenter, M. P. (1977). The distribution of world science. *Social Sudies of Sience,* 7(4), 501-516.

Hood, C., & Peters, B. G. (2004). The Middle Aging of New Public Management: Into the Age of Paradox? *Journal of Public Administration and Theory,* 14(3), 267-282.

Laudel, G., & Origgi, G. (2006). Introduction to a special issue on the assessment of interdisciplinary research. *Research Evaluation,* 15(1), 2-4.

Leydesdorff, L. (2008). *Caveats* for the Use of Citation Indicators in Research and Journal Evaluation. *Journal of the American Society for Information Science and Technology,* 59(2), 278-287.

Leydesdorff, L., & Meyer, M. (2010). The Decline of University Patenting and the End of the Bayh-Dole Effect, *Scientometrics* (in print); DOI: 10.1007/s11192-009-0001-6.

Leydesdorff, L., & Wagner, C. S. (2008). International collaboration in science and the formation of a core group. *Journal of Informetrics,* 2(4), 317-325.

Leydesdorff, L., & Wagner, C. S. (2009). Is the United States losing ground in science? A global perspective on the world science system. *Scientometrics,* 78(1), 23-36.

Martin, B. R. (2009, forthcoming). Inside the public scientific system: changing modes of knowledge production. In S. Kuhlmann, P. Shapira & R. Smits (Eds.), *Innovation Policy: Theory and Practice, an International Handbook.* Cheltenhalm, UK: Edward Elgar.

Merton, R. K. (1968). The Matthew Effect in Science. *Science,* 159, 56-63.

Münch, R. (2008). Stratifikation durch Evaluation. Mechanismen der Konstruktion von Statushierarchien in der Forschung. *Zeitschrift für Soziologie*, 37, 60-80.

Osterloh, M., & Frey, B. S. (2008). *Creativity and conformity: Alternatives to the present peer review system*. Paper presented at the workshop on *Peer Review Reviewed*. Berlin: WZB, 24-25 April 2008.

Persson, O., & Melin, G. (1996). Equalization, growth and integration of science. *Scientometrics,* 37(1), 153-157.

Plomp, R. (1990). The significance of the number of highly cited papers as an indicator of scientific prolificacy. *Scientometrics,* 19(3), 185-197.

Power, M. (2005). The theory of the audit explosion. In E. Ferlie, L. E. Lynn jr. & C. Pollit (Eds.), *The Oxford Handbook of Public Management* (pp. 326-344). Oxford: Oxford University Press.

Price, D. de Solla (1976). A general theory of bibliometric and other cumulative advantage processes. *Journal of the American Society for Information Science,* 27(4), 292–306.

Rafols, I., & Leydesdorff, L. (2009). Content-based and Algorithmic Classifications of Journals: Perspectives on the Dynamics of Scientific Communication and Indexer Effects *Journal of the American Society for Information Science and Technology,* forthcoming.

Ritzer, G. (1998). *The McDonaldization Thesis*. London: Sage.

Rousseau, R. (1992). *Concentration and diversity measures in informetric research*. Ph.D. Thesis, University of Antwerp, Antwerp.

Rousseau, R. (2001). Concentration and evenness measures as macro-level scientometric indicators (in Chinese). In G.-h. Jiang (Ed.), *Research and university evaluation*

*(Ke yan ping jia yu da xue ping jia)* (pp. 72-89). Beijing: Red Flag Publishing House.

Sala-i-Martin, X. (2006). The world distribution of income: falling poverty and ... covergence, period. *Quarterly Journal of Economics,* 71(2), 351-397.

Schmoch, U., & Schubert, T. (2009). Sustainability of incentives for excellent research - the German case. *Scientometrics*, 81(1), 195-218.

Shimank, U. (2005). New public management and the academic profession: reflections on the German situation. *Minerva,* 43, 361-376.

Stiftel, B., Rukmana, D., & Alam, B. (2004). Faculty quality at US graduate planning schools: A National Research Council-style study. *Journal of Planning Education and Research,* 24(1), 6-22.

Timothy, M. S. (2005). Public Policy, Economic Inequality, and Poverty: The United States in Comparative Perspective. *Social Science Quarterly,* 86(s1), 955-983.

Van den Besselaar, P., & Leydesdorff, L. (2009). Past performance, peer review, and project selection: A case study in the social and behavioral sciences. *Research Evaluation,* 18(4), 273-288.

Van Parijs, P. (2009). European Higher Education under the Spell of University Rankings. *Ethical Perspectives,* 16(2), 189-206.

Van Raan, A. F. J. (2005). Fatal Attraction: Conceptual and methodological problems in the ranking of universities by bibiometric methods. *Scientometrics,* 62(1), 133-143.

Ville, S., Valadkhani, A., & O'Brien, M. (2006). Distribution of Research Performance Across Australian Universities, 1992-2003, and its Implications for Building Diversity. *Australian Economic Papers,* 45(4), 343-361.

Weingart, P. (2005). Impact of Bibliometrics upon the science system: Inadverted Consequences? *Scientometrics,* 62(1), 117-131.

Weingart, P., & Maasen, S. (2007). Elite through rankings - The emergence of the enterprising university. In R. Whitley & J. Gläser (Eds.), *The Changing Governance of the Sciences: The Advent of Reserch Evaluation Systems* (pp. 75-99). Dordrecht: Springer.

Whitley, R. D. (1984). *The Intellectual and Social Organization of the Sciences*. Oxford: Oxford University Press.

Zitt, M., Barré, R., Sigogneau, A., & Laville, F. (1999). Territorial concentration and evolution of science and technology activities in the European Union: a descriptive analysis. *Research Policy,* 28(5), 545-562.